\documentclass[final,5p,times,twocolumn]{elsarticle}

\usepackage{amssymb}
\usepackage{epsfig}
\usepackage{epstopdf}
\usepackage[breaklinks=true,colorlinks=true,linkcolor=blue,urlcolor=blue,citecolor=blue]{hyperref}
\usepackage{cite}
\usepackage{lineno}

\begin{document}

\begin{frontmatter}

\title{An alternative for teaching and learning the simple diffusion process using Algodoo animations}

\author{Samir L. da Silva$^{a}$$^{, *}$; Judismar T. Guaitolini Junior$^{a}$;\\
 Rodrigo L. da Silva$^{b}$; Emilson. R. Viana$^{c}$ and F\'{a}bio F. Leal$^{d}$}

\address{$^{a}$Instituto Federal do Esp\'{i}rito Santo, campus Vit\'{o}ria, 29040-780 Vit\'{o}ria, Brazil.\\
$^{b}$Instituto Federal Fluminense, campus Bom Jesus do Itabapoana, 28360-000 Bom Jesus do Itabapoana, Brazil.\\
$^{c}$Universidade Tecnol\'{o}gica Federal do Paran\'{a}, campus Curitiba, DAFIS, 80230-901. Curitiba, Brazil.\\
$^{d}$Instituto Federal Fluminense, campus Campos Centro, 28030-130 Campos dos Goytacazes, Brazil.\\
\vspace{0.5pc}
$^{*}$samir.lacerda@ifes.edu.br}

\begin{abstract}

In this work animations of the random walk movement using a freeware Algodoo were done in order to support teaching the concepts of Brownian Motion. The random walk movement were simulate considering elastic collision between the particles in suspension in a fluid, and the particles which constitute the fluid. The intensity of velocities where defined in an arbitrary range, and we have a random distribution of the velocity directions. 
Using two methods, the distribution histogram of displacements (DHD) and the mean-square-displacement ${\langle{\Delta r^{2}}\rangle}$ (MSD), it was possible to measure the diffusion coefficient of the system, and determine the regions where the system presents ballistic regime or diffusive transport regime. The ballistic regime was observed graphically when the MSD has a parabolic dependence with time, which differing from the typical diffusive regime where MSD has a linear dependence. The didactical strategy for combining analytical approaches as graphic analysis, and animations in software’s with easy implementation supports the teaching and learning processes, especially in Physics were we want to explain experimental results within theoretical models.

\end{abstract}

\begin{keyword}

Brownian motion; Random Walk; Diffusion Coefficient; Animation; Algodoo; Teaching of Physics.

\end{keyword}

\end{frontmatter}

\section{Introduction} 
\label{sec1}

The diffusion is one of the most studied and spread processes in science. The Einstein's description about the erratic motion of small particles on fluid surfaces, the historical Brownian Motion (BM) ~\cite{r01,r02,r03,r04}, lead to a proof that matter is constituted by atoms and molecules in constant motion in according to the kinetic-theory. Other numerous physical systems exhibit some diffusion process as the spreading of dengue by migration of infected individual or mosquitoes \cite{r05}, the diffusion-limited aggregation applied to growth phenomena \cite{r06}, the analysis of financial data on stock market \cite{r07}, the model of protein folding \cite{r08}, the fixational eye movements \cite{r09} and many more. Many of these stochastic processes can be studied using the classical random walk (RW). 

It is not easy to obtain an intuitive comprehension of the stochastic phenomena, because this usually requires advanced mathematical tools. Actually, several teachers presenting a good pedagogical approach by using numerical experiments of low computational cost, and that are available to students with access to a computer. Also, many other numeric-experiments can be used to complement real experiences. The use of computational technologies is explored by many educators that propose strategies to elaborate good tools for supporting the cognitive development of the students. Such technologies have been widely used in the physics teaching \cite{r10,r11,r12}. To minimize the difficulties of teaching and learning  physical phenomena, some teachers has added to your classes some softwares able to determine the evolution of the equations, in order to create simulations and animations of present phenomena, or even, to automatization of the experimental data acquisition, to modeling and interacting with virtual physical environments \cite{r13,r14,r15,r16,r17,r18,r19}.

Therefore, in this paper, the simulation of the two-dimensional random walk movement of particles on fluid surface, in order to study the Brownian motion, was done by using computational animations. To accomplish this task, we use the free software for 2D simulation known as Algodoo of Algoryx Simulation AB$\textsuperscript{\textregistered}$\cite{r20}. This software has an interactive environment, allowing creation of experimental scenarios, as animated movies, but with having as feedback, the equations and physical properties imposed to the simulations. The Algodoo is an easy manipulation tool, and does not require specific knowledge about computer programming or training to realize tasks with the software, which allow easy learning by students. In a recent work of our research group \cite{r21} we present the potentiality of that software as a tool for teaching and learning physics by considering the launching projectiles animations.

We believe that the complementation of mathematical description of the Brownian motion by using the animations with Algodoo will provide to the students a better learning about all techniques involved in this stochastic model. With this project we can also create the basis for the theoretical instrumentation for studying others diffusive process when we use the Algodoo's animations for the Brownian motion.  

This paper is organized as follows; firstly, we will introduce some of the basic properties of random walks in two-dimension for an active particle with random direction velocity in a two-dimensional homogeneous environment. For convenience, we will use the familiar picture of one diffusing particle. Then the animations were built by using the Algodoo software. Next, from the animations, we will show how to calculate the diffusion coefficient by using two methods: the mean-square-displacement and the displacement histogram of the Brownian particle. And finally, we show that the random walk animations provide a clear understanding of transition between the ballistic behavior and the diffusive behavior of Brownian motion.

\section{The Brownian motion and the procedure for building animations}
\label{sec2}

The Brownian system is constituted by a suspended particle in a fluid with random motion resulting for their collision with the atoms and molecules of the fluid. Since, is not an easily task to simulate a fluid, because this simulation involve to determine the solutions of the Euler and Navier-Stokes equations. So, in our animations the fluid is formed by small blue disks moving randomly.

The manufacture of this fluid consist in to create a little set of identical disks and to attribute to them velocity with different magnitude and direction randomly. To increase the number of disks of the fluid the process are repeated several times until we get a desired concentration. The null friction and maximum elasticity in the collision between the disks is considered, to maintain constant the mean kinetic energy of system. These characteristics can be established selecting all disks and to edit in the item ``\textit{Material}" of the Algodoo software. A red disk can be introduced in any region of the system, and represents our “Brownian” particle. The particles of the fluid (red disks) has a diameter defined much bigger than the particles in suspension (blue disks) in order to enhance the contrast. 

For the present animation was estabilished, by convinience, a flat rectangular region of ${575.0\;}$ $m^{2}$, where were uniformly distributed 7.2$\times$$10^3$ blue disks forming the system fluid. Each blue disk has a mass of ${25\;}$ g and occupies an area of 12.0$\times$$10^{-2}$${\;}$${m^2}$. Intensities of velocities were distributed between 0.1 m/s e 5.5 m/s with random directions. The Brownian particles, i.e., the red disk, has a mass of 70 g and area of 3.0$\times$$10^{-1}$${\;}$ ${m^2}$. These specification were choose arbitrary and do not represent any material or specific system. This software has a limitation to assign small dimensions and mass in the drawn objects. Our aim is to create fictitious animations of the random walk movement to study the Brownian motion properties.

\begin{figure}[h]
\begin{center}
\includegraphics[width=1.0\linewidth]{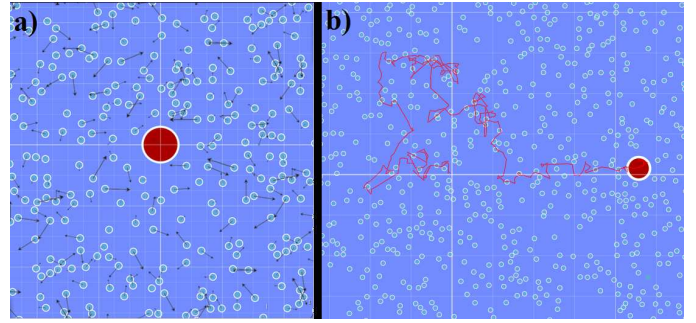}
\end{center}
\caption{(Color online) Illustration of Brownian motion animation. The small discs represent the fluid and the larger disk is a Brownian particle that will perform the randomic motion because of collisions between the particles of the fluid.(See video submitted)}
\label{Figure1}
\end{figure}

In the fig. 1 (a) we present the illustration of the environment built to observe the Brownian motion. When we start the animations in the Algodoo environment, the simulate dynamics of the blue disks a random motion of the molecules in a fluid (fig. 1 (b)). To see trajectories of the red disk we use the tool ``\textit{Tracer}". This tool draws a line by all the trajectory of the selected object. The trajectory of the motion can be observed in the fig. 1 (b) by a path illustrated in a red line. 

Theoretically, the motion in the \textit{x} and \textit{y} axes obeys the classical random walk with the two-dimensional mean displacement $\langle{\Delta r}\rangle$ null and the mean square displacement (MSD) $\langle{\Delta r^{2}}\rangle$ is given by:

\begin{equation}
 \langle{\Delta r^{2}}\rangle = \langle{\Delta x^{2}}\rangle + \langle{\Delta y^{2}}\rangle = 4Dt
.\label{eq01}
\end{equation}

\vspace*{2mm}
\noindent
where $\langle{\Delta x^{2}}\rangle$ and $\langle{\Delta y^{2}}\rangle$ are the one-dimensional MSD, \textit{D} is the diffusion coefficient and \textit{t} the time between the points of the data \cite{r01,r02,r03,r04,r22,r23}. Thus, the MSD depends of the diffusion coefficient and time. The root mean square (RMS) displacement $\langle{\Delta r^{2}}\rangle$$_{rms}$= $\sqrt{\langle{\Delta r^{2}}\rangle}$ is proportional to $\sqrt{t}$, allowing that the red disk visits quickly \textit{(t$\downarrow$)} the surrounding area of the starting point and that requires a long times \textit{(t$\uparrow$)} to get long distances \cite{r23}.

The BM presents a transition between a smooth ballistic behavior to the diffusive behavior. The scaling-time of the transition is given by the relaxation time $\tau$. In real physical systems, the relaxation time $\tau$ is ordinarily very short, typically in the order of some microseconds or nanoseconds ~\cite{r23,r24,r25,r26}. For a short interval of time ${(\Delta t<\tau}$), ${\langle{\Delta r^{2}}\rangle}$ has a quadratic time dependence, and for long times (${\Delta t\gg\tau}$) this dependence becomes linear \cite{r23,r24,r25,r26}. Even the theory was proposed in 1905 \cite{r02}, the transition between ballistic and diffusive behavior was experimentally proofed only in 2010 \cite{r24}, when was possible to measure the position of the particle to study the instantaneous velocity and the transition of the ballistic to the diffusive behavior. 

\begin{figure}[h]
\begin{center}
\includegraphics[width=1.0\linewidth]{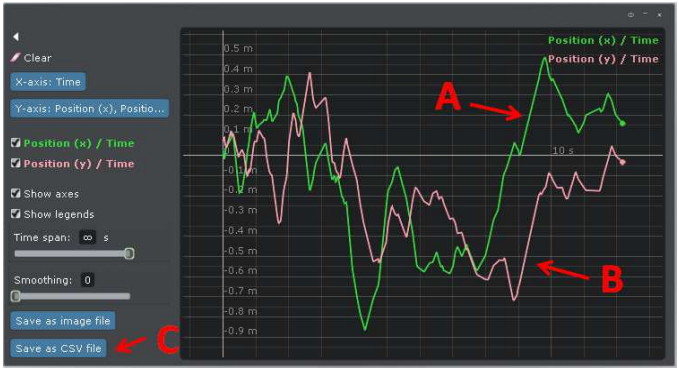}
\end{center}
\caption{(Color online) Algodoo graph which presents the time series for \textit{x} and \textit{y} positions for Brownian disc.}
\label{Figure2}
\end{figure}

To study the BM by animation with Algodoo we use the graphical tool ``\textit{Show Plot}". With this tool we can chose one of the variables used in the animation and present them in a graphic, as illustrated in the fig. 2. In the case of BM, the time evolution of the horizontal \textit{x(t)} and the vertical \textit{y(t)} positions of the Brownian disk, was treated as time series. The option ``\textit{Show Plot}" exhibits the \textit{x} and \textit{y} positions of the disk as a function of time, that can be observed in the fig. 2 items (\textit{A}) and (\textit{B}) respectively. With this tool is possible to save a spreadsheet in an extension .CSV (item (\textit{C}) of the fig. 2) with a frequency of 60 points per second. Was used one hundred Brownian disks and monitored the time evoluting of each disk during ten seconds. Since any disk presents a different initial condition in the one hundred CSV files were produced containing the time series of the Brownian disks, and the time series were used to analyze the BM.

In our animations, the diffusion coefficient \textit{D} obtained from two methods. The first method is the evaluation of the graphic of the probability-distribution of the particle position. Using this graphic we can fit the gaussian curve by using:

\begin{equation}
P(x) = \frac{1}{{\sigma \sqrt {2\pi } }}e^{{{ - \left( {x - \langle{\Delta x}\rangle } \right)^2 } \mathord{\left/ {\vphantom {{ - \left( {x - \mu } \right)^2 } {2\sigma ^2 }}} \right. \kern-\nulldelimiterspace} {2\sigma ^2 }}}
.\label{eq03}
\end{equation}
\vspace*{2mm}

\noindent
the variance $\sigma^2$ in the Brownian motion is equal to the MSD ${\langle{\Delta x^{2}}\rangle}$, since the mean displacement ${\langle{\Delta x}\rangle}$ is null in this case.  

\begin{equation}
 \sigma^2 = \langle{\Delta x^{2}}\rangle - \langle{\Delta x}\rangle^2
.\label{eq04}
\end{equation}
\vspace*{2mm}

We can also use the variance $\sigma^2$ from the displacements distribution ${\langle{\Delta x^{2}}\rangle}$ to calculate \textit{D} by equation of one-dimensional random walk:

\begin{equation}
 \sigma^2 = \langle{\Delta x^{2}}\rangle =2Dt
.\label{eq05}
\end{equation}
\vspace*{2mm}

Using any software that analyze mathematically the spreadsheet, as the Microsoft Excell, Open Office or Origin, we can do the mathematical treatment of the time series, as desired. By calculating the horizontals and verticals displacements $\Delta x_i=x_i(t)-x_i(t_0)$ and $\Delta y_i=y_i(t)-y_i(t_0)$ it is possible to create a list and to produce the graphic of the probability distribution of the particle displacements, which will be fitted using equation 3.

The second method consists in calculate the two-dimensional MSD for each instant of time t by:

\begin{equation}
 \langle{\Delta x^{2}(t)}\rangle = \frac{1}{N} \sum_{i=1}^{N} \left[\left(x_i(t)-x_0(t_0)\right)^2+\left(y_i(t)-y_0(t_0)\right)^2\right]
.\label{eq06}
\end{equation}

\vspace*{2mm}
\noindent
where the cartesian pair ${\left(x_i(t_0),y_i(t_0)\right)}$ represents the initial position of the particle in each sample and \textit{N} is the number of samples. The diffusion coefficient \textit{D} is the slope of the linear fit of the curve ${\langle{\Delta x^{2}(t)}\rangle}$ versus \textit{t}, according to equation 1.

To demonstrate the transition between the ballistic and diffusive behavior we use the MSD curve for a short interval of time, generated by the second method. From the graphic of $\langle{\Delta r^{2}(t)}\rangle$ versus \textit{t} we can demonstrate the quadratic and linear temporal behavior of $\langle{\Delta r^{2}(t)}\rangle$ and from this graphic the relaxation time $\tau$ is obtained.

In laboratories classes for undergraduate students, an experiment about a typical BM requires that the student register the motion of at least ten particles for measure the diffusion coefficient and/or the relaxation time. Experiments of BM in advanced stages became significant, but the cost of preparation and the quantification of these experiments are difficult, usually producing deviations about 10$\%$ to 15$\%$ or even more \cite{r22}.

In practical terms, the study of the Brownian motion by computational experiments minimize largely the experimental problems. In this scenario, the Algodoo software is a good option and it stand out by easier manipulation, and does not requiring a specific programming knowledge. We do not propose here the elimination of BM experiments, but we present a new tool to support the teaching/learning of physics by animations built with Algodoo.

\section{Results and Discussions}
\label{sec3}

\begin{figure}[hbt]
\begin{center}
\includegraphics[width=1.0\linewidth]{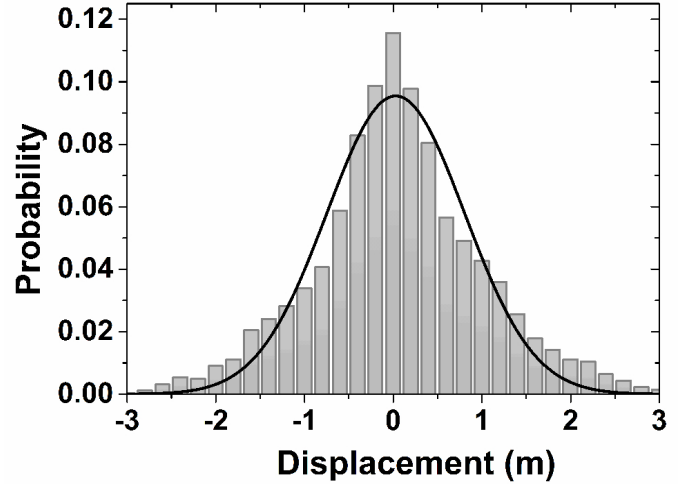}
\end{center}
\caption{In method 1, the diffusion coefficient is found by making a histogram of the \textit{x} and \textit{y} displacements (gray) and a Gaussian fit (black line) for the data; the variance is used to calculate the diffusion coefficient \textit{D} = 0.221 $m^2$/s.}
\label{Figure3}
\end{figure}

With the animations done, teachers can establish with their students a great opportunity for discussions about the excitement of the small disks associating qualitatively to thermal excitement of the hypothetical fluid. Such thermal excitement produces continuous collisions with the red disk located in that region, causing an erratic motion, characterizing the Brownian motion (fig. 1 (a) e (b)). Due to the fact that this software can be easy manipulate, teachers can pause, increase or decrease the speed of the animations, or also modify the environment characteristic to enrich the qualitative presentation of the motion.

As said in previous section, the time series of the Brownian disk positions is saved, by ``\textit{Show Plot}" tool, in an extension .CSV where data are analyzed by worksheet software. We have executed one hundred independently Brownian motions for 10 s. During this interval of time we have exclude all initial points till 1.5 s. Until approximately 0.8 s we observe that the animations do not get the random walk state and consequently it can disturb the analysis of the motion. These observations are done for the animation with the initial conditions described in the previous section. If those conditions change, can one realize new analysis of the initial time interval. Therefore, we have established a 1.5 s, for convenience.

By using method $\#$1, we have fitted a Gaussian curve (equation 3) on the displacement distribution graphics in the fig. 3. The histogram is built from the independently one-dimensional displacements $\Delta x$ and $\Delta y$ together. The variance of the adjusted Gaussian curve is equal to the mean-square of one-dimensional displacement $\langle{\Delta x^{2}}\rangle$, and we can use this to calculate the diffusion coefficient by using equation 5. The obtained value from the displacement histogram is \textit{D} = 0.221 m${^2}$/s.

\begin{figure}[h!]
\begin{center}
\includegraphics[width=1.0\linewidth]{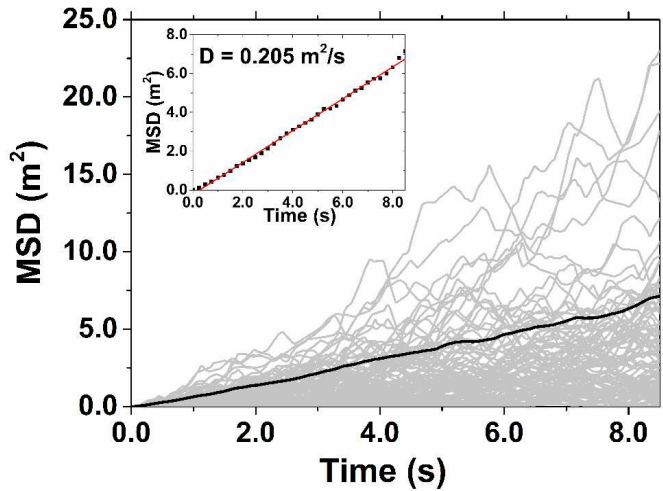}
\end{center}
\caption{(Color online) (Gray) The square displacement of 100 samples. (Black) Mean square displacement versus time. Inside the linear fit for MSD with \textit{D} = 0.205 $m^2$/s.}
\label{Figure3}
\end{figure}

After the square position displacements were calculated for all Brownian disks in the animation (fig. 4 – grey curves), we calculate the average over all samples, i.e., MSD (fig. 4 – black curve), by equation 6. The diffusion coefficient is determined by the slope of MSD as function of time by using equation 1, where we obtain \textit{D} = 0.205 m${^2}$/s (using method $\#$2). We found a difference of about 8$\%$ between the methods.

For a short time interval (${\Delta t < \tau}$) the dynamics of the Brownian disk is governed by the translacional inertia and the motion is in ballistic regime. Is demonstrated in figure 5 that MSD has a parabolic time dependence (curve with red points) differing from the typical diffusive motion (black points), where MSD has a linear dependence. The relaxation time was availed at $\tau$ = 0.15 s. We have adjusted a curve with the form $\langle{\Delta r^{2}}\rangle$(t) = C$t^2$+Bt+A on the parabolic region and we obtain  C = 1.9 $m^2$/$s^2$, B = 3.5${\times}$${10^{-2}}$ ${m^2}$/s and A = -1.9${\times}$${10^{-4}}$ ${m^2}$.

Changing the animation of the BM it's possible to generate others diffusion regimes, that can be used to teach other diffusive phenomena, as electrical diffusive motion on semiconductor materials.

\begin{figure}[hbt]
\begin{center}
\includegraphics[width=1.0\linewidth]{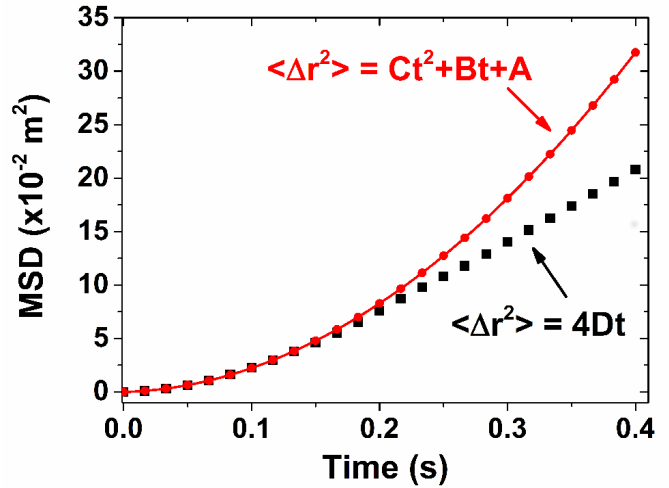}
\end{center}
\caption{(Color online) MSD is ballistic for (${\Delta t < \tau}$) with $\langle{\Delta r^{2}}\rangle$ $\propto$ $t^2$ - Adjust (red curve) with . (Black) For long times (${\Delta t \gg \tau}$) $\langle{\Delta r^{2}}\rangle$ becomes linear. The relaxation time was estimated at $\tau$ = 0.15 s.}
\label{Figure5}
\end{figure}

These animations can be used in several levels in the school. For the high school level, the animations of the BM  can be used to illustrate the random motion of a particle suspended on a fluid. For undergraduate levels, the animations can be also a tool to facilitate comprehension of mathematics procedures involved in the BM studying. From the data collected by Algodoo, the teacher can elaborate scripts to teach the students how to obtain physical quantities from the time series, the same way as is done in laboratory experiments. Also, it would be an interesting introduction to the concepts of statistical physics, normal distribution, Gaussian curve, and so on which is very important in many areas of physics, from thermodynamics to quantum mechanics.

For higher levels, students can learn scientific procedures for formulation of models, collecting and data treatment. This basic knowledge supports students, beginning in the science world, by development of research projects.

\section{Conclusions}
\label{sec4}

In this paper, we showed a computational tool of great potential for teaching and learning physics, using the freeware Algodoo. Using an animation based on the two-dimensional random walk movement of particles on fluid surface we can study the basic concepts of the Brownian motion as a great support-tool for teaching/learning of statistical physics.

The animations presented in this paper provide to teachers and students a simple tool for quantitative and quantitative analyzes of the BM. It is possible to observe and to discuss the the random walk movement of the small disks resulting in an erratic motion, and to associate ones qualitatively to the thermal excitement of a hypothetical fluid, which characterize the BM.

The diffusion coefficient was calculated from the animations by two methods: Method $\#$1 – fitting the graphic of the distribution of the independently one-dimensional displacements $\Delta x$ and $\Delta y$ together by a Gaussian curve. And Method $\#$2 – by the fitting the graphic MSD as function of the time, using equation 1. It was found a difference of approximately 8$\%$ between these two methods.

We have demonstrated that for a short interval of time (${\Delta t < \tau}$), $\langle{\Delta r^{2}}\rangle$ is proportional to \textit{t${^2}$} and the motion presents a ballistic behavior, were we have a movement of the particle without suffering collisions in the path. For long times ${\Delta t \gg \tau}$, $\langle{\Delta r^{2}}\rangle$ is proportional to \textit{t}, and the motion is diffusive, movement suffering collisions along the path. The time of relaxation was availed as $\tau$ = 0.15 s.

The Algodoo is easy to manipulate and does not require any specific knowledge in programming or training to realize the tasks in the software. Through this environment, educators and students can explore all potentialities of the studied theme, proposing modifications: in the size of the particles, in the intensity of the velocities, imposing boundary conditions; to the system and discussing their consequences: enhance of the relaxation time, reduction of the diffusive coefficient, and so on.

Many diffusive systems can be explored by these initial proposal of an animation of the Brownian motion. The didactic strategy for combining analytical approaches and animations, supports teaching and learning processes. In this way Algodoo also show one as a support-tool \cite{r20} that becomes the teaching/learning process of physics simpler and fruitful when compared to others software of animations and learning \cite{r10,r11,r12,r13,r14,r15,r16,r17,r18,r19}.

\section{Acknowledgments}
\label{sec4}

The authors thanks FAPERJ for the finantial support of this work.


\end{document}